\documentclass[%
nofootinbib,
 amsmath,amssymb,
 aps, 
 physrev,
]{revtex4-2}

\usepackage{graphicx}
\usepackage{dcolumn}
\usepackage{bm}
\usepackage{xcolor}

\begin{document}


\title{\textbf{High-Power Targetry for Muon Production\\Contribution to the 25th International Workshop on Neutrinos from Accelerators} 
}%

\author{Michael T Hedges}
\email{Corresponding author: mhedges@fnal.gov}
\affiliation{%
 Fermi National Accelerator Laboratory, Batavia, IL, USA.
}%

\author{Madeleine Bloomer}
\affiliation{
Department of Physics, Emory University, Atlanta, GA, USA.
}%



\begin{abstract}
The production of high-intensity muon beams is crucial for advancing particle and accelerator physics, both now and in the future. Achieving these high-intensity goals requires overcoming significant challenges in high-power targetry. Here, we will outline the key challenges and explore the selection of target materials, focusing on muon production and susceptibility to radiation damage---key factors for optimal target designs.

\end{abstract}

\maketitle


\section{\label{sec:intro}Introduction}

The experimental sensitivities of many current and next-generation high-energy particle physics experiments depend on the availability of high-intensity secondary beams, such as muon beams and neutrino beams. These beams are typically produced by directing a high-energy proton beam onto a target material, initiating inelastic nuclear interactions that generate pions. The pions subsequently decay, yielding both muons and neutrinos. While neutrino production has been a well-established area of research, muon production has only recently regained significant attention due to its pivotal role in fundamental physics studies and its potential for applications in future accelerator facilities. Near-term muon experiments, such as Mu2e \cite{mu2e-su2020}, which aims to search for charged lepton flavor violation (CLFV), will soon require the highest intensity muon beams yet achieved. Furthermore, the proposed high-energy muon collider \cite{mucol-2024}, for instance, promises to open new avenues for precision measurements and particle interactions, offering unique advantages over traditional electron-positron or proton-proton colliders. However, realizing these ambitious experimental programs will require advancements in high-power targetry. This will include careful selection of target materials and designs that can withstand intense irradiation damage, manage the resulting thermal and mechanical stresses, and maintain a high level of muon production.

Generating a high-intensity muon beam requires efficient capture of produced muons, necessitating the use of a large solenoid with a very strong magnetic field,\footnote{For context, the upcoming Mu2e Experiment will use a production solenoid operating at a maximum field strength of 4.6 T, and muon collider studies suggest a solenoidal field strength of 20 T.}, which guides the muons toward a beamline. In this scheme, the production target must be compact enough to fit within the confines of the solenoid. Therefore, a muon production target must achieve all three of the following criteria: it must maximize muon production, it must be long-lived under proton beam irradiation, and it must be compact.

Current state-of-the-art solutions in high-power targetry, dominated by needs for production of neutrino and neutron beams which do not require high field extraction solenoids, have shared only two of these goals: maximizing secondary beam intensity and maintaining target survivability in megawatt class proton beams. 
They have used large-mass targets to mitigate heat, shock load, and material damage caused by the primary beam. As examples, the neutrino target in the Neutrinos at the Main Injector (NuMI) beamline at Fermilab utilizes a $1.2~\mathrm{m}$ long graphite target with a proton beam energy of $120~\mathrm{GeV}/c^2$, and the first target station at the Spallation Neutron Source (SNS) at Oak Ridge National Laboratory utilizes many liters of flowing liquid mercury with a proton beam energy of $\sim{}1~\mathrm{GeV}/c^2$.

In contrast, Mu2e requires a target designed to fit within a compact cylindrical volume of just 3 mm in radius and 220 mm in length. This compactness presents a significant challenge, as the target is more susceptible to radiation damage due to its limited mass. The target's small size also limits its ability to spread out thermal and mechanical loads, furthering the need for careful choice of target material to maintain performance and prevent failure under prolonged beam irradiation.

\section{Radiation Damage}
Radiation damage is a key factor in target survivability as it alters the physical properties of the target material, including thermal conductivity, mechanical strength, and shock resistance. High-energy proton beams induce defects like dislocations and vacancies, and the effects of fragmentation and transmutation further reduce target performance. The extent of radiation damage is often quantified using displacements per atom (DPA), which summarizes the degree of radiation-induced damage to the lattice. DPA can be calculated per incident proton using simulation tools like FLUKA \cite{fluka-cern,fluka-og}, which model the interactions between the proton beam and the target material. As DPA increases, the material's ability to manage heat and absorb shock diminishes, accelerating degradation and failure.

\begin{figure}[bh]
\includegraphics[width=0.8\textwidth{}]{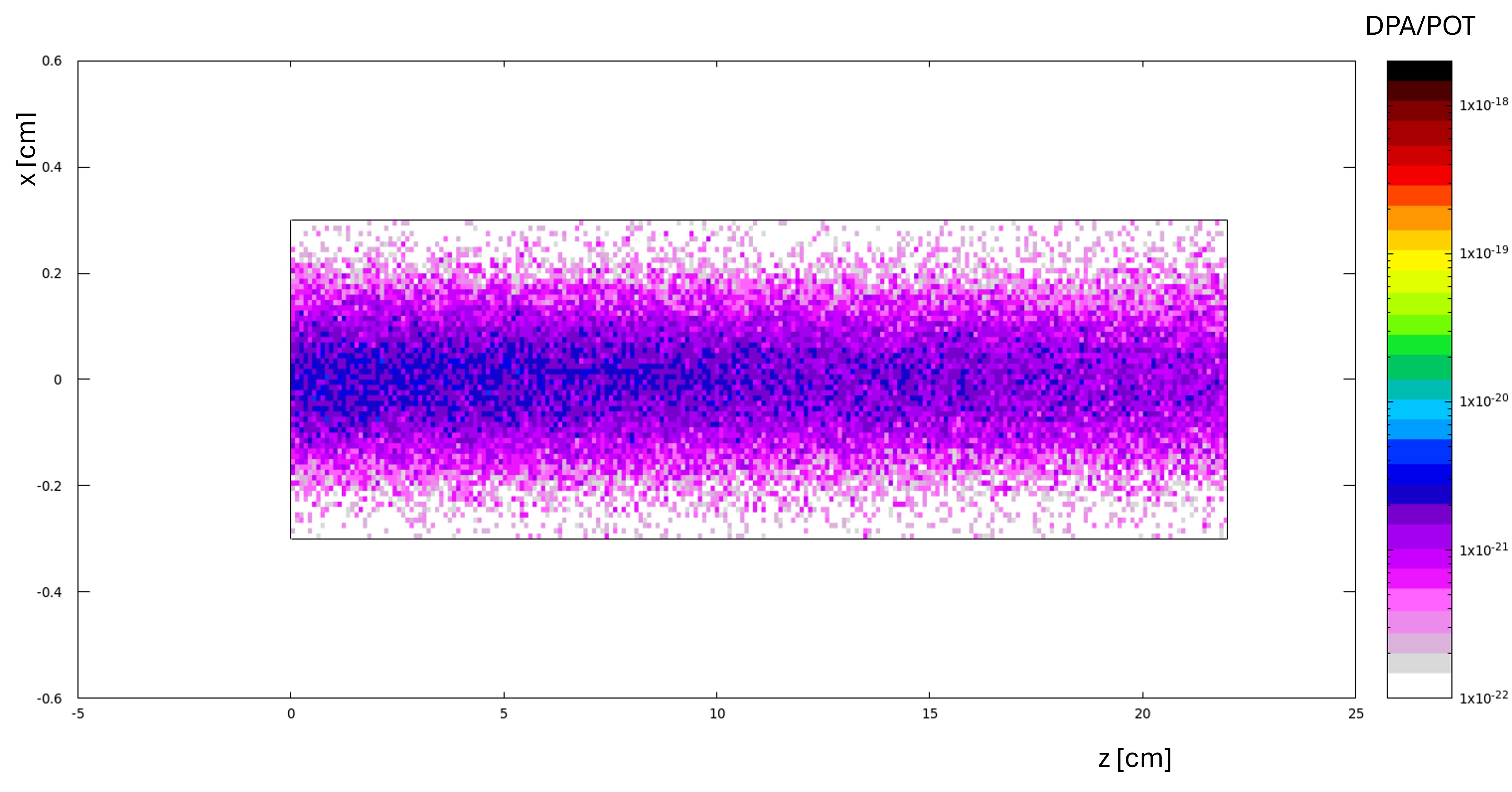}
\includegraphics[width=0.8\textwidth{}]{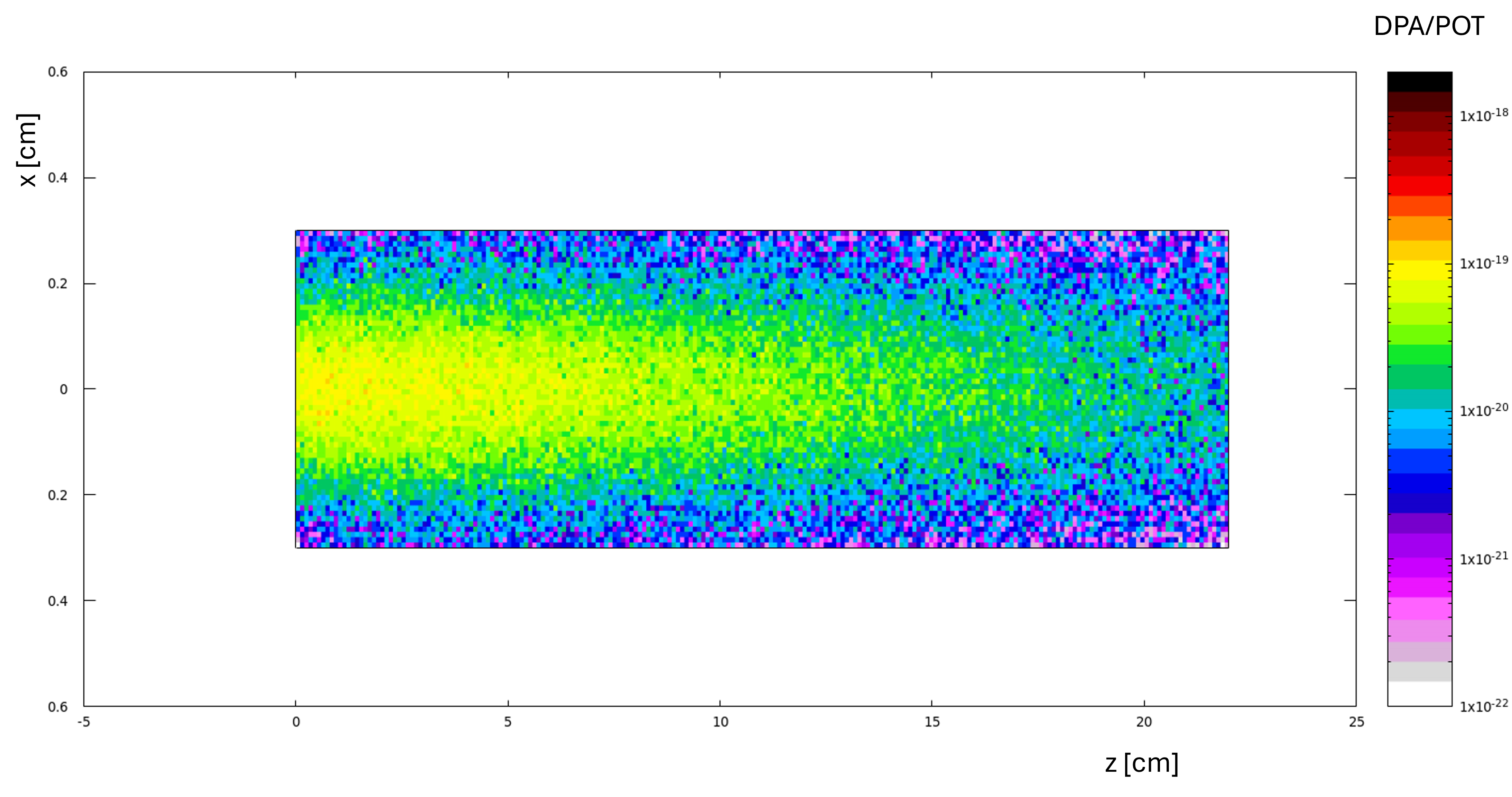}
\caption{\label{fig:graphInc} FLUKA simulation of NRT-dpa per proton on target within the central $|y| < 0.01$ cm slice for graphite (top) and Inconel-600 (bottom) targets subjected to the Mu2e proton beam. Incoming proton beam starts at $(0,0,0)$ and moves along $\hat{z}$.}
\end{figure}

\begin{figure}[th]
\includegraphics[width=0.8\textwidth{}]{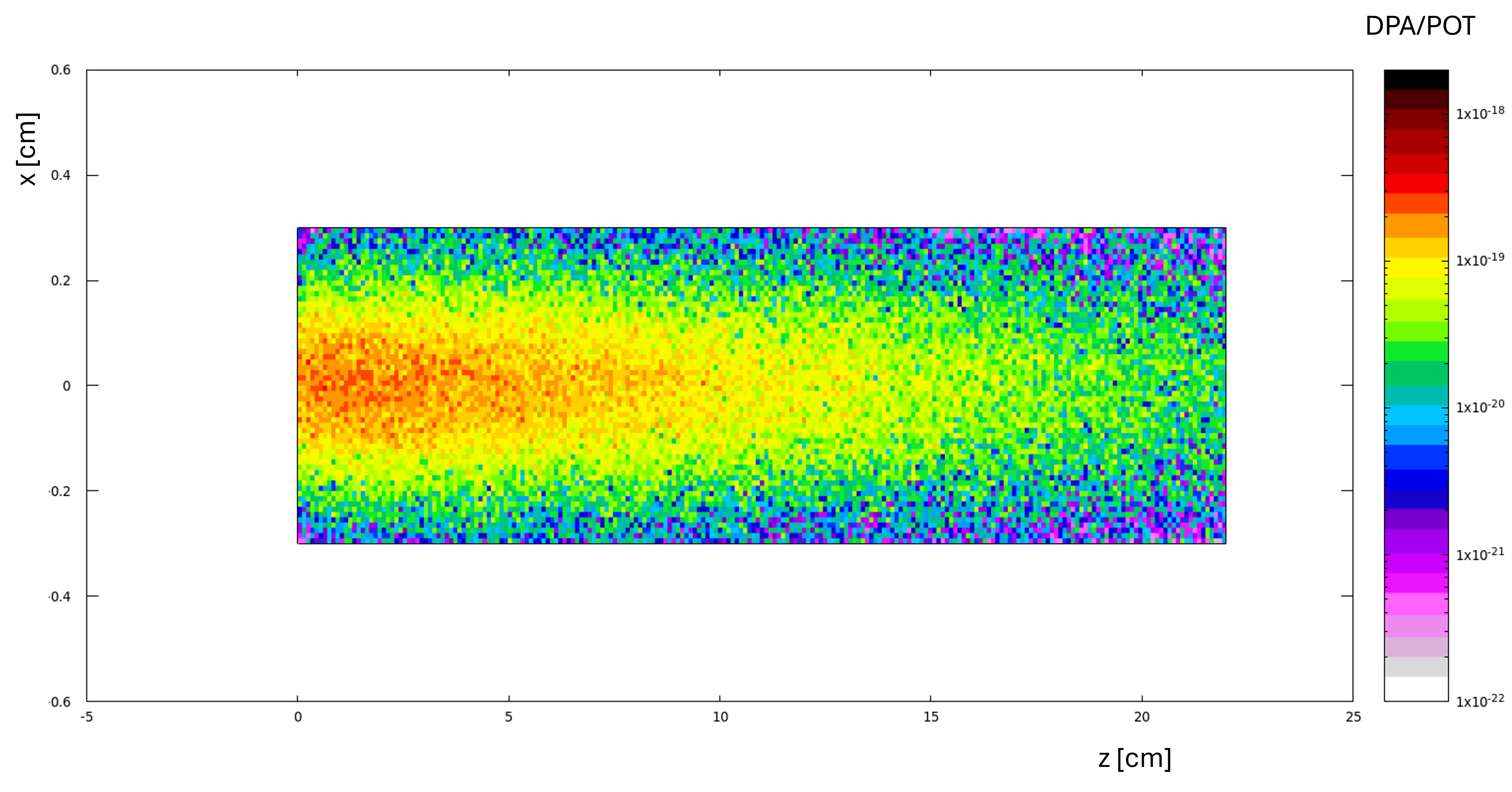}
\includegraphics[width=0.8\textwidth{}]{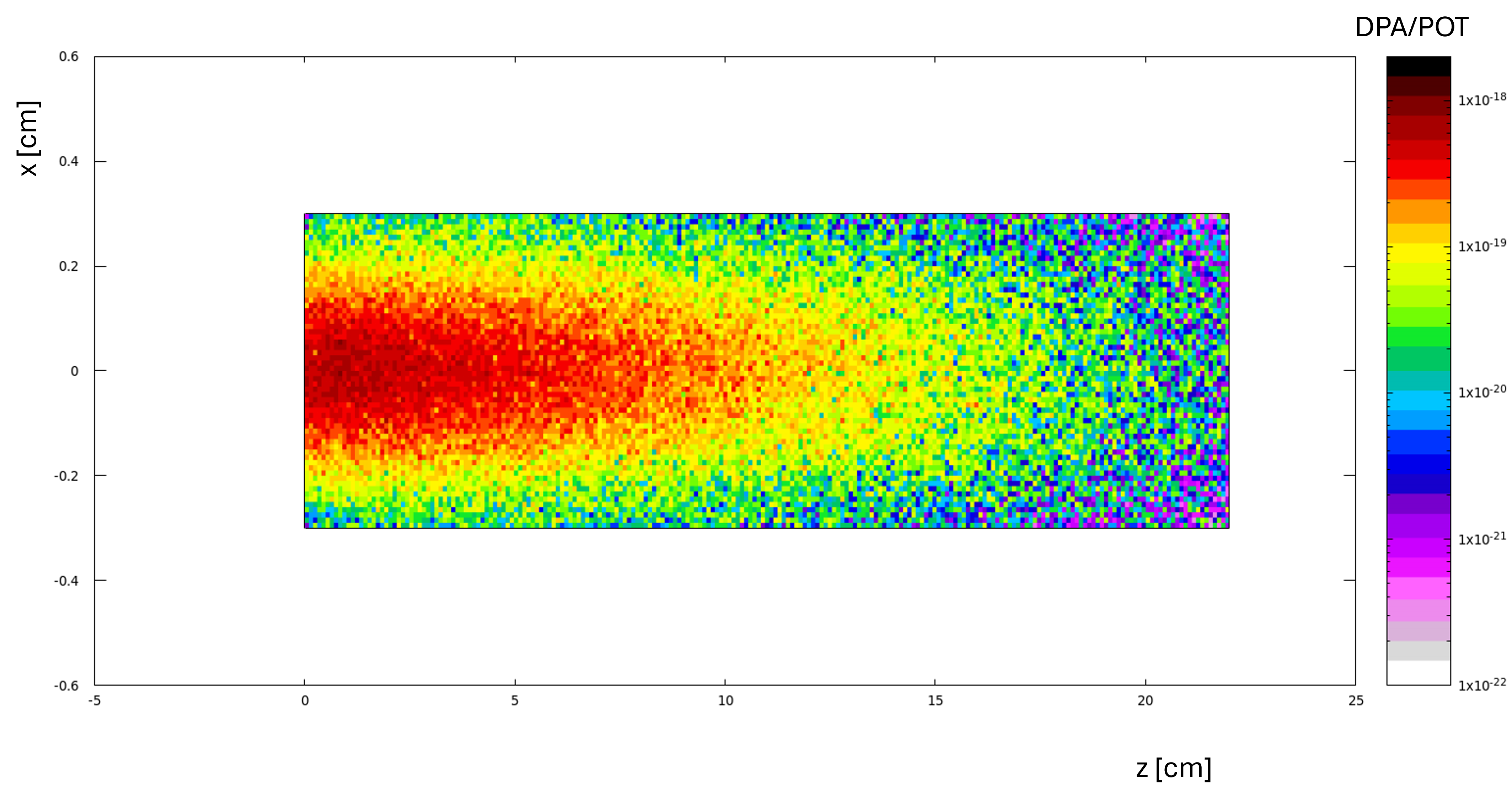}
\caption{\label{fig:TZM-W} FLUKA simulation of NRT-dpa per proton on target within the central $|y| < 0.01$ cm slice for TZM (top) and tungsten (bottom) targets subjected to the Mu2e proton beam. Incoming proton beam starts at $(0,0,0)$ and moves along $\hat{z}$.}
\end{figure}

Figs. \ref{fig:graphInc} and \ref{fig:TZM-W} show FLUKA calculations of the Norgett-Robinson-Torrens DPA (NRT-dpa) per proton on target (POT) for 3 mm radius $\times$ 220 mm length cylindrical targets made of graphite, Inconel-600, Titanium-Zirconioum -Molybdenum (TZM), and pure tungsten. The simulated proton beam matches design parameters of the Mu2e beam of $\sigma=1~\rm{mm}$ radius at 8 GeV/$c^2$ beam energy. While not an exhaustive set, these materials represent realistic target material choices at various densities. All figures share the same normalization and color scale, peaking at just over $1\times10^{-18}$ NRT-dpa per POT in the tungsten target. Assuming a nominal operational target replacement frequency of once per year during scheduled summer shutdowns, the design beam delivery of $\sim{}1\times10^{20}$ POT/year would induce as little as a manageable $O(1)$ peak DPA in graphite and up to a never-before-seen $O(100)$ peak DPA in tungsten. While further study is needed to draw complete conclusions, these findings show an extreme increase in DPA/POT with increasing density of the target material.

\section{Muon Production}
\begin{figure}[h]
\includegraphics[scale=0.8]{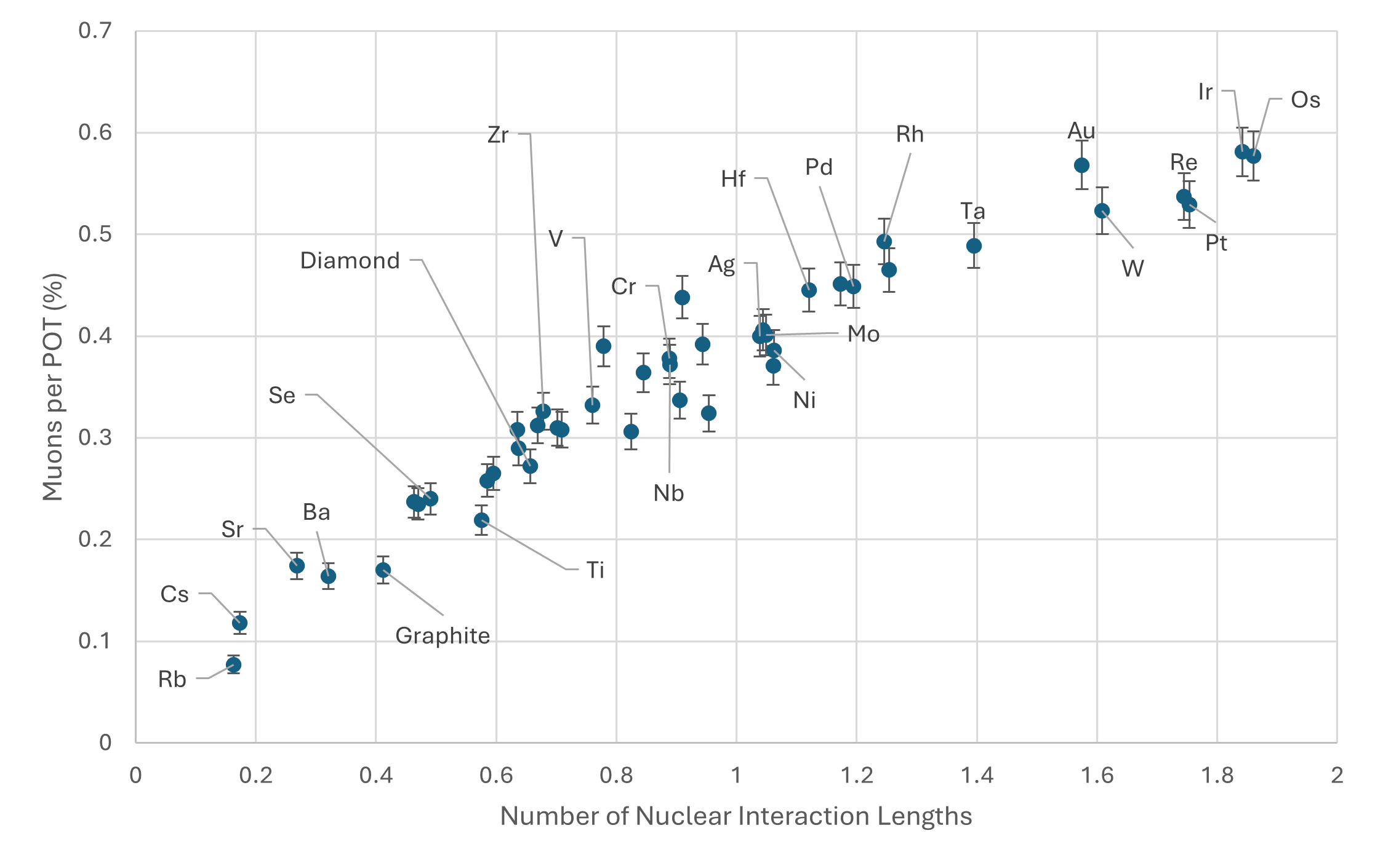}
\caption{\label{fig:mu-vs-lambda} Muon production per proton on target for the Mu2e beam as a function of the number of nuclear interaction lengths in a 220 mm long cylindrical target of various materials, simulated using G4Beamline (adapted from Ref. \cite{osti_2426573}).}
\end{figure}

The choice for target material also impacts the muon yield per incident proton. The common line of reasoning has been that highest density targets must result in the highest levels of muon production. This has led to a preference in proposed target designs for materials such as mercury, tantalum, tungsten, and even iridium. With the same target geometry and Mu2e beam conditions, we used G4Beamline \cite{g4beamline} to calculate the number of produced muons in various target materials. In Fig. \ref{fig:mu-vs-lambda}, we find significant diminishing returns in muon production with higher density targets. Muon production is roughly linearly dependent on the number of nuclear interaction lengths in the target, which does not increase linearly with target density.\footnote{For more details, we refer the reader to the study done in Ref. \cite{osti_2426573}.}

\section{Summary and Discussion}
The production of high-intensity muon beams from primary proton beams in current and next-generation experiments presents a unique challenge: the need for compact targets that fit within an extraction solenoid. In this exploratory analysis, we highlight that radiation damage can quickly become a limiting factor for the performance and lifetime of compact, solid, non-rotating, high-density targets. Radiation damage, quantified by DPA, scales superlinearly with increasing target density, while muon production scales with the number of nuclear interaction lengtths and increases only marginally with density. This disparity strongly motivates further investigation into material optimization, particularly for medium-density materials like nickel-based Inconel, used in the Fermilab Antiproton Source \cite{Morgan:2003zza}, and molybdenum-based alloys such as TZM, which has seen limited study in high-power targetry. The use of optimized materials, coupled with strategies like active cooling and rotational mechanisms, may be key to meeting the demanding requirements of future muon production facilities.

\section{Acknowledgments}
This manuscript has been authored by Fermi Research
Alliance, LLC under Contract No. DE-AC02-07CH11359
with the U.S. Department of Energy, Office of Science,
Office of High Energy Physics.

\nocite{*}

\bibliography{nufact2024}

\end{document}